# Ultrasmall Glutathione-Protected Gold Nanoclusters as Next Generation Radiotherapy Sensitizers with High Tumor Uptake and High Renal Clearance

Xiao-Dong Zhang,[a] Zhentao Luo,[b] Jie Chen,[a] Shasha Song,[a] Xun Yuan,[b] Xiu Shen,[a] Hao Wang,[a] Yuanming Sun,[a] Kai Gao,[c] Lianfeng Zhang,[c] SaijunFan,[a] David Tai Leong,[b] Meili Guo,[d*] and Jianping Xie [b*]

[a] Tianjin Key Laboratory of Radiation Medicine and Molecular Nuclear Medicine, Institute of Radiation Medicine, Chinese Academy of Medical Sciences and Peking Union Medical College, No. 238, Baidi Road, Tianjin, 300192, China

[b] Department of Chemical and Biomolecular Engineering, National University of Singapore, 4 Engineering Drive 4, Singapore 117585, Singapore

[c] Key Laboratory of Human Disease Comparative Medicine, Ministry of Health, Institute of Laboratory Animal Science, Chinese Academy of Medical Sciences & Comparative Medical Center, Peking Union Medical College, Beijing, 100021, China

[d] Department of Physics, School of Science, Tianjin Chengjian University, Tianjin 300384, China

**Abstract:** Radiotherapy is often the most straightforward first line cancer treatment for solid tumors. While it is highly effective against tumors, there is also collateral damage to healthy proximal tissues especially with high doses. The use of radiosensitizers is an effective way to boost the killing efficacy of radiotherapy against the tumor while drastically limiting the received dose and reducing the possible damage to normal tissues. Here, we report the design and application of a good radiosensitizer by using ultrasmall $Au_{29-43}(SG)_{27-37}$ nanoclusters (<2 nm) with a naturally-occurring peptide (e.g., glutathione or GSH) as the protecting shell. The GSH-coated $Au_{29-43}(SG)_{27-37}$ nanoclusters can escape the RES absorption, leading to a good tumor uptake (~8.1% ID/g at 24 h post injection). As a result, the as-designed Au nanoclusters led to a strong enhancement for radiotherapy, as well as a negligible damage to normal tissues. After the treatment, the ultrasmall $Au_{29-43}(SG)_{27-37}$ nanoclusters can be efficiently cleared by the kidney, thereby avoiding potential long-term side-effects caused by the accumulation of gold atoms in the body. Our data suggest that the ultrasmall peptide-protected Au nanoclusters are a promising radiosensitizer for cancer radiotherapy.

Cancer remains one of the world's most devastating diseases with more than 10 million new cases each year, and radiotherapy is a leading cancer treatment approach that addresses the needs of more than 50% cancer patients[1]. Though high-energy radiation can fatally damage tumor cells, it can also harm normal tissues. In fact, the mitotically active tumor cells are only slightly more susceptible to radiation damage than those in the essential normal tissues[2]. Hence, it is very important to strike the right balance between eradicating tumor and saving normal tissues by controlling the target and the dose of radiation administered to the patient. Many improvements have been made in radiotherapy to target tumors better, which could cause less damage to normal tissues. For example, megavolt (6–25 MV) X-rays are now used to avoid skin damage; tomotherapy and intensity-modulated radiation therapy (IMRT) are applied to better concentrate the radiation within the tumor volume; and optimal dose fractionation schedules are also developed to allow better cumulative damages to the tumor and adequate repairing of normal tissues[2-5]. Despite such advances, it is still challenging to use radiotherapy alone to eradicate tumor cells. A magic bullet to current challenges in radiotherapy is radiosensitizer, which can locally increase the efficacy of radiotherapy by enhancing the radiation damages to the cell.

In general, the radiosensitizing agents can be classified into two major categories according to their mechanisms of action: (type-1) chemotherapeutics that modulate the cell response to enhance the radiation damage, and (type-2) materials that interact directly with the radiation and generate additional damages to the cell[2,3,6,7]. The development of type-1 radiosensitizers started with Heidelberger's preclinical studies[8] in 1958, and this radiosensitizing approach is often referred to as combined chemotherapy and radiotherapy or chemoradiation[3,9]. Most organic radiosensitizers are type-1, which enhance radiotherapy by modulating cell responses, such as reducing the radioresistance of tumor cells, preventing the formation of blood vessels (or disrupting the existing vessels, anti-angiogenic), inducing apoptosis, and suppressing mitosis[3,6,8]. Although many preclinical and clinical studies have affirmed the efficacy of type-1 radiosensitizers, a major drawback of these chemotherapeutics is their inherent cytotoxicity and side effects. For example, gemcitabine is known to cause myelosuppression, anemia, vomiting, and diarrhea[10,11]. Similarly, cisplatin is known to have myelotoxicity, neurotoxicity, and nephrotoxicity, and it can also cause hemolytic anemia, hearing loss, and vomiting[12-14].

Type-2 radiosensitizers are mostly metal-based materials that can strongly absorb, scatter, and reemit radiation energy, resulting in a local radiation dose increase when they are accumulated in tumors[15,16]. Intense research on nanoscale metallic materials in the past two decades has provided many novel materials for biomedical applications[17]. Among these emerging radiosensitizers, gold nanoparticles (Au NPs) are particularly attractive because of their strong interaction with the radiation (Au has a high atomic number of 79), excellent chemical stability and inertness, and good biocompatibility (low toxicity)[18-21].

The enhancement of radiation dose received by the tumor tissue loaded with Au relative to the dose received by normal tissues without Au can be 200% or higher[22,23]. Such enhancement comes from the direct interaction between Au and radiation. When the incident radiation (gamma rays, X-rays) impinges on a Au NPs, the NPs becomes a new source of radiation and emits high energy through scattered photons (X-rays), photoelectrons, Compton electrons, Auger electrons, electron–positron pairs, and fluorescence photons, thereby causing radiochemical (free radicals and ionization) damages to the surrounding tumor tissue[22,24,25]. However, most of the Au NPs that have been demonstrated so far have large particle sizes (typically above 50 nm) and could be trapped by the reticulo-endothelial system (RES) absorption, which could result in low tumor uptake and unavoidable accumulation in liver and spleen[26-31]. Decreasing the particle size could benefit the escape of particles from the RES absorption. For example, one recent study showed that Au NPs with particle sizes below 20 nm could efficiently escape the RES absorption and showed good tumor uptake[32]. However, the sizes of these particles were still above the renal clearance barrier, that is ~5.5 nm, and could therefore induce the accumulation of NPs in RES, thus resulting in potential toxicity over the long term[33-36]. Besides the core size of NPs, the protecting ligands on the NPs surface can also affect the *in vivo* biodistribution. For example, the naked Au NPs of particle sizes of 1.9 and 4.8 nm, while small, have low colloidal stability due to the protein corona acquired in blood. These Au NPs eventually formed large aggregates of ~20-100 nm, which could not be rapidly metabolized and certainly unable to escape the RES[18,37]. Au NPs with different surface ligands can induce different NPs-protein corona in blood that could determine the RES absorption and cellular uptake efficiency[38].

Taken together of the two key attributes (size and surface) for NPs-based radiosensitizers, we hypothesized that: 1) small naturally-occurring peptides, such as glutathione or GSH, could be a good surface ligand for Au NPs by helping them escape the RES absorption and improving their deposition in tumors; and 2) ultrasmall Au NPs with core sizes below 2 nm (hereafter referred to as nanoclusters, NCs) in combination with the GSH ligands can ensure a small hydrodynamic diameter (HD), which could provide good interface with the biological system, improve their *in vivo* pharmacokinetics, and enhance their deposition in tumors[39]. Here we demonstrate such concept by using sub-2-nm GSH-protected Au NCs with a well-defined molecular formula of $Au_{29-43}(SG)_{27-37}$[40]. We show in this study that the $Au_{29-43}(SG)_{27-37}$ NCs have attractive features of high tumor uptake, strong sensitizing enhancement for radiation, and low toxicity, and they could be a good candidate for next generation radiosensitizers for clinical use. This study has therefore enriched the family of Au NPs and NCs that could show good performance for cancer radiotherapy[33,37].

**Results and Discussion**

The Au$_{29-43}$(SG)$_{27-37}$ NCs were prepared by a reported procedure[40]. The as-prepared Au NCs showed a shoulder peak at ~400 nm in the UV-vis absorption spectrum (Figure 1a), and surface plasmon resonance (SPR, typically at ~520 nm, a characteristic absorption of large Au NPs) was not observed. The molecular-like absorption of these Au NCs could be attributed to the discrete electronic states arising from the ultrasmall size of the NCs[40-45]. A representative transmission electron microscopy (TEM, Figure 1b) image confirmed that the Au NC cores were smaller than 2 nm. The hydrodynamic diameter (HD) of Au$_{29-43}$(SG)$_{27-37}$ NCs was determined to be ~2.8 nm by using dynamic light scattering (DLS, Figure 1c). In addition, Au$_{29-43}$(SG)$_{27-37}$ NCs showed strong orange luminescence with an emission peak at ~610 nm (Figure 1d, black line), which was also consistent with the previous report[40].

We tested the blood stability of the as-prepared Au$_{29-43}$(SG)$_{27-37}$ NCs and the extent of plasma protein that binds to the NCs by missing Au$_{29-43}$(SG)$_{27-37}$ NCs (0.5 mL, 3 mM per Au atom) with blood plasma (0.5 mL). The photoluminescence of the mixture of Au$_{29-43}$(SG)$_{27-37}$ NCs and blood plasma (at 24 h after mixing) was not decreased significantly as compared with the aqueous solution of the NCs (Figure 1d), suggesting that Au$_{29-43}$(SG)$_{27-37}$ NCs were sufficiently stable in blood. The unbound Au$_{29-43}$(SG)$_{27-37}$ NCs were separated from the protein-bound Au NCs by filtering the mixture of Au$_{29-43}$(SG)$_{27-37}$ NCs and blood plasma (at 24 h after mixing) using ultrafiltration with a molecular weight cut-off, MWCO of 50 kDa. About 40% of Au$_{29-43}$(SG)$_{27-37}$ NCs were recovered from the filtrate as determined by their photoluminescence intensity (Figure S1), indicating that the binding ratio of plasma protein was ~60%.

We further performed *in vivo* experiments to investigate the pharmacokinetics of the Au$_{29-43}$(SG)$_{27-37}$ NCs. The mice were intraperitoneally injected with the Au$_{29-43}$(SG)$_{27-37}$ NCs (~5.9 mg-Au/kg-body). As shown in Figure 2a, the distribution half-life (first phase $t_{1/2\alpha}$) of Au$_{29-43}$(SG)$_{27-37}$ NCs in blood was determined to be 6.5 h. As compared with the reported Au$_{10-12}$(SG)$_{10-12}$ and Au$_{25}$(SG)$_{18}$ NCs, the longer distribution half-life of the Au$_{29-43}$(SG)$_{27-37}$ NCs could be attributed to their larger hydrodynamic diameters[33,46]. The concentration of Au$_{29-43}$(SG)$_{27-37}$ NCs in blood was gradually stabilized after ~12 h (Figure 2a). The high concentration of Au$_{29-43}$(SG)$_{27-37}$ NCs in blood may lead to high tumor uptake of the NCs.

The tumor uptake of the Au$_{29-43}$(SG)$_{27-37}$ NCs was measured using inductively coupled plasma mass spectrometry (ICP-MS, Figure 2b). The tumor uptake of the Au NCs reached a maximum at 24 h post injection (p.i.), corresponding to 8.1% ID/g (9.5 μg/g). The tumor uptake gradually decreased from 24 to 48 h p.i. The observed tumor uptake was higher than that of the previously reported PEG-coated Au nanorods (~7.1% ID/g)[27], Au NPs (~3% ID/g)[29,37], small Au NCs (~2.3–3.2% ID/g)[47]. We recently reported two kind of clusters, Au$_{25}$(SG)$_{18}$ and Au$_{10-12}$(SG)$_{10-12}$, and their tumor uptake were determined to be 13% and 50% ID/g, respectively[33,46]. In general, smaller particles may feature with higher tumor uptake. Compared with Au$_{25}$(SG)$_{18}$ and Au$_{10-12}$(SG)$_{10-12}$, the tumor uptake of Au$_{29-43}$(SG)$_{27-37}$ is relatively lower. However, one salient point of Au$_{29-43}$(SG)$_{27-37}$ is its strong orange emission at 610 nm with a high quantum yield of 15%; such strong emission could be advantageous for some biomedical applications. The ratios of the concentration of Au in tumor relative to that in other tissues and organs are important parameters to evaluate the specificity of the NCs. The tumor/kidney, tumor/blood, and tumor/liver ratios were determined to be 2.1/1.0, 4.5/1.0, and 14.2/1.0, respectively.

Detailed biodistribution and clearance of $Au_{29-43}(SG)_{27-37}$ NCs were further investigated. Figure 2c shows the biodistributions of $Au_{29-43}(SG)_{27-37}$ NCs at 24 h and 28 days p.i. Tumor and kidney possessed predominant distributions relative to spleen, liver, heart, and lung at 24 h p.i., which supports that $Au_{29-43}(SG)_{27-37}$ NCs could escape RES absorption and achieve efficient targeting. The majority of Au were cleared at 28 days p.i. because only 0.2% ID/g Au in liver, ~0.4% ID/g Au in kidney, and <0.1% ID/g in tumor were found, suggesting a high efficacy of renal clearance of Au NCs[48,49]. In contrast, many other inorganic nanomaterials, such as Au NPs, carbon nanotubes, and graphene, are difficult to be cleared[28,37,50,51]. It is worth mentioning that the $Au_{29-43}(SG)_{27-37}$ NCs with GSH ligands on the NC surface featured with a different biodistribution from that of the Cy5-labeled $Au_{25}(SG)_{18}$[46]. The possible reason could be the Cy5 labeling, which might modify the surface chemistry of $Au_{25}(SG)_{18}$[52]. However, in the pristine $Au_{29-43}(SG)_{27-37}$ NCs, the GSH ligand on the NC surface may help mitigate the serum protein adsorption[53].

We also confirmed the tumor uptake and efficient renal clearance of $Au_{29-43}(SG)_{27-37}$ NCs by the X-ray computed tomography *in vivo* imaging (Figure 3). X-ray CT imaging is a non-invasive and reliable method for tumor imaging. The CT signal depends on the concentration of Au in tissues. A CT value of 1212 HU corresponding to 60 mM of Au (Figure S2), which is a good value for *in vivo* imaging. In this study, the as-prepared $Au_{29-43}(SG)_{27-37}$ NCs (60 mM Au, 0.2 mL) were injected into mice *via* tail vein, and two- and three-dimensional X-ray CT images were recorded. We measured the tumor uptake of $Au_{29-43}(SG)_{27-37}$ NCs using U14 tumor bearing mice. As shown in Figure 3a and 3b, the corresponding CT value was determined to be 365 HU, which was much higher than that of the muscle tissue (214 HU). A significant tumor uptake was observed in the tumor site (indicated by the arrows, Figure 3a) at 6 h p.i. In addition, a clear boundary between tumor and normal tissue was observed. Figure 3c and 3d showed the renal clearance of $Au_{29-43}(SG)_{27-37}$ NCs at the time points of 1 and 24 h p.i., measured using nude mice without tumor. The bladder (indicated by the arrow, Figure 3c) showed high contrast at 1 h p.i. (1300 HU), and this value (383 HU) was obviously decreased at 24 h p.i., indicating the efficient clearance of $Au_{29-43}(SG)_{27-37}$ NCs by kidney[49].

We also examined the cancer radiation treatment of $Au_{29-43}(SG)_{27-37}$ NCs by using U14 tumor bearing nude mice as the animal model. The mice were intraperitoneally injected with $Au_{29-43}(SG)_{27-37}$ NCs of a concentration of 5.9 mg-Au/kg-body. As a maximum tumor uptake of $Au_{29-43}(SG)_{27-37}$ NCs was reached at 24 h p.i. (Figure 2b), the mice were irradiated under $^{137}$Cs gamma radiation of 3600 Ci at a 5 Gy dose at 24 h p.i. At 28 days p.i., the tumor volumes and weights in the sacrificed mice were measured (Figure 4a). Compared with the control group, a remarkable decrease (~76%) of tumor volume was observed in mice treated with Au$_{29-43}$(SG)$_{27-37}$ NCs plus radiation (p<0.05). In addition, compared with the mice treated by radiation only, the tumor volume decreased to ~66% in mice treated with Au$_{29-43}$(SG)$_{27-37}$ NCs plus radiation (p<0.05). Figure 4b showed that the tumor weight decreased in mice treated with Au$_{29-43}$(SG)$_{27-37}$ NCs plus radiation. Similarly, a significant tumor weight decrease was seen in mice treated with Au$_{29-43}$(SG)$_{27-37}$ NCs plus radiation relative to that in mice treated with radiation only, suggesting that the Au$_{29-43}$(SG)$_{27-37}$ NCs can enhance the radiation therapy.

We finally checked the toxicological responses by examining blood biochemistry (Figure 5) and pathology (Figure 6) of the mice. No significant weight loss, drastic organ or blood chemistry changes were found, suggesting that the renal clearable Au$_{29-43}$(SG)$_{27-37}$ NCs did not induce a significant liver and kidney toxicity. In contrast, the naked Au NPs, PEG-coated Au NPs, and BSA-protected Au NCs with the hydrodynamic diameter of ~6-100 nm have been found with acute liver toxicity, such as the increase of alanine aminotransferase (ALT) and aspartate aminotransferase (AST)[37,50,54,55,56]. Traditional radiosensitizers, such as cisplatin, also showed high kidney toxicity due to slow clearance[57]. Thus, the Au$_{29-43}$(SG)$_{27-37}$ NCs developed in this study could emerge as an attractive radiosensitizing agent with its low toxicity and high tumor uptake.

In summary, the Au$_{29-43}$(SG)$_{27-37}$ NCs covered by GSH can escape the RES absorption and showed high tumor accumulation *via* the improved EPR effect. The hydrodynamically ultrasmall Au$_{29-43}$(SG)$_{27-37}$ NCs showed very efficient renal clearance, and no obvious toxicity was observed in the body. The as-designed Au NCs can also significantly enhance the efficacy of the cancer radiotherapy. These advantageous features allow the Au$_{29-43}$(SG)$_{27-37}$ NCs to be attractive radiosensitizer materials for further testing.

**Methods**

**Synthesis and Characterizations of Au$_{29-43}$(SG)$_{27-37}$ NCs.** The synthesis and purification of Au$_{29-43}$(SG)$_{27-37}$ NCs followed the published procedures[40,58]. Briefly, freshly prepared aqueous solutions of HAuCl$_4$ (20 mM, 0.50 mL) and GSH (100 mM, 0.15 mL) were mixed with 4.35 mL of ultrapure water at 25 °C. The reaction mixture was heated to 70 °C under gentle stirring (500 rpm) for 24 h. An aqueous solution of intensely orange-emitting Au$_{29-43}$(SG)$_{27-37}$ NCs was formed. The orange-emitting Au$_{29-43}$(SG)$_{27-37}$ NC solution could be stored at 4 °C for 6 months with negligible changes in their optical properties. The as-prepared Au$_{29-43}$(SG)$_{27-37}$ NCs were purified through ultrafiltration (3 kDa membrane).

*In vivo* **Biodistribution.** The studies were approved by the Institute of Radiation Medicine, Chinese Academy of Medical Sciences and Animal Care Research Advisory Committee of Institute of Radiation Medicine, Chinese Academy of Medical Sciences, while experiments conducted following the guidelines of the Animal Research Ethics Board of Institute of Radiation Medicine, Chinese Academy of Medical Sciences. Forty-eight mice were purchased, maintained, and handled using protocols approved by the Institute of Radiation Medicine, Chinese Academy of Medical Sciences (CAMS). The U14 tumor models were generated by subcutaneous injection of 2 × 10$^6$ cells suspended in 50 μL of PBS into the right shoulder of male nude mice. The mice treated with Au$_{29-43}$(SG)$_{27-37}$ NCs were sacrificed at 0.5, 1, 2, 6, 12, 24, 48, and 72 h post injection (p.i.). The main organs, such as tumor, liver, kidney, spleen, heart, lung, brain were collected. The organs of Au$_{29-43}$(SG)$_{27-37}$ NCs treated mice were digested using a microwave system CEM Mars 5 (CEM, Kamp Lintfort, Germany) to determine their Au content, which was determined by an inductively coupled plasma mass spectrometer (Agilent 7500 CE, Agilent Technologies, Waldbronn, Germany).

*In vivo* **Imaging.** Eighteen mice were purchased, maintained, and handled using protocols approved by the Institute of Radiation Medicine, Chinese Academy of Medical Sciences (CAMS). The U14 tumor models were generated by subcutaneous injection of $2 \times 10^6$ cells suspended in 50 μL of PBS into the right shoulder of male nude mice. Before the experiments, the mice were anesthetized by chloral hydrate. For CT imaging, 200 μL of GSH-protected $Au_{29-43}(SG)_{27-37}$ NCs (60 mM, 0.2 mL) were injected through the intraperitoneal routes into mice. Each mouse was imaged on a small-animal scanner (microPET/CT, Inveon, Siemens). The mice were exposed to a 10-min CT scan and the images were reconstructed using the filtered back-projection algorithm with CT-based photon-attenuation correction. CT data were analyzed for regions of interest, including tumor, bladder, and spleen.

*In vivo* **Radiation Therapy.** All animals were purchased, maintained, and handled using protocols approved by the Institute of Radiation Medicine, CAMS. The U14 tumor models were generated by subcutaneous injection of $2 \times 10^6$ cells suspended in 50 μL of PBS into the right shoulder of BALB/c mice. The male mice were intraperitoneally treated with the $Au_{29-43}(SG)_{27-37}$ NCs when the tumor volume reached 100–120 mm$^3$ (7 days after tumor inoculation). For each treatment, $Au_{29-43}(SG)_{27-37}$ NCs (0.59 mg-Au/mL) were intraperitoneally injected at a dosage of 5.9 mg/kg in the mice. As the control, 200 μL of saline was intraperitoneally injected into each mouse in the control group. Subsequently, the mice were irradiated by 5 Gy gamma-rays from $^{137}$Cs (photon energy 662 keV) with an activity of 3600 Ci at 24 h p.i. for $Au_{29-43}(SG)_{27-37}$ NCs injections. Thirty two male mice were assigned to the following four groups (eight mice per group): control, $Au_{29-43}(SG)_{27-37}$, radiation alone, and $Au_{29-43}(SG)_{27-37}$ + radiation. The tumor size was measured every two or three days, and calculated using the equation: tumor volume = (tumor length) × (tumor width)$^2$/2.

*In vivo* **Toxicity.** The treated mice were weighed and assessed for behavioral changes. All mice were sacrificed at 28 days p.i., and their blood and organs were collected for hematology, biochemistry and toxicological investigation. The blood was drawn for hematology analysis (potassium EDTA collection tube) and serum biochemistry analysis (lithium heparin collection tube) using a standard saphenous vein blood collection technique. During necropsy, liver, kidney, spleen, heart, lung, brain, genitals, tumor, and thyroid were collected and weighed. Major organs from these mice were then fixed in 4% neutral buffered formalin, processed into paraffin, and stained with hematoxylin and eosin (H&E). Pathology was examined using a digital light microscope.

**References**

1. Jemal, A. *et al.* Global cancer statistics. *CA-Cancer J. Clin.* **61**, 69-90 (2011).
2. Brown, J. M. & Workman, P. Partition Coefficient as a guide to the development of radiosensitizers which are less toxic than misonidazole. *Radiat. Res.* **82**, 171-190 (1980).
3. Wardman, P. Chemical radiosensitizers for use in radiotherapy. *Clin. Oncol.* **19**, 397-417 (2007).
4. Jain, S. *et al.* Cell-specific radiosensitization by gold nanoparticles at megavoltage radiation energies. *Int. J. Radiat. Oncol. Biol. Phys.* **79**, 531-539 (2011).
5. Yasui, H. *et al.* Radiosensitization of tumor cells through endoplasmic reticulum stress induced by PEGylated nanogel containing gold nanoparticles. *Cancer lett.* **347**, 151-158 (2014).
6. Kasid, U. & Dritschilo, A. RAF antisense oligonucleotide as a tumor radiosensitizer. *Oncogene* **22**, 5876-5884 (2003).
7. Kvols, L. K. Radiation Sensitizers: A selective review of molecules targeting DNA and non-DNA targets. *J. Nucl. Med.* **46**, 187S-190S (2005).
8. Heidelberger, C. *et al.* Studies on fluorinated pyrimidines: II. effects on transplanted tumors. *Cancer Res.* **18**, 305-317 (1958).
9. Herskovic, A. *et al.* Combined chemotherapy and radiotherapy compared with radiotherapy alone in patients with cancer of the esophagus. *New. Engl. J. Med.* **326**, 1593-1598 (1992).
10. Robson, M. *et al.* Quality of life in women at risk for ovarian cancer who have undergone risk-reducing oophorectomy. *Gynecol. Oncol.* **89**, 281-287 (2003).
11. Aapro, M. S., Martin, C. & Hatty, S. Gemcitabine-a safety review. *Anti-Cancer Drugs* **9**, 191-202 (1998).
12. Legha, S. S. & Dimery, I. W. High-dose cisplatin administration without hypertonic saline: observation of disabling neurotoxicity. *J. Clin. Oncol.* **3**, 1373-1378 (1985).
13. Bokemeyer, C. *et al.* Analysis of risk factors for cisplatin-induced ototoxicity in patients with testicular cancer. *Br. J. Cancer* **77**, 1355-1362 (1998).
14. Carozzi, V. *et al.* Effect of the chronic combined administration of cisplatin and paclitaxel in a rat model of peripheral neurotoxicity. *Eur. J. Cancer* **45**, 656-665 (2009).
15. Ali, H. & van Lier, J. E. Metal complexes as photo- and radiosensitizers. *Chem. Rev.* **99**, 2379-2450 (1999).

16  Butterworth, K. T., McMahon, S. J., Currell, F. J. & Prise, K. M. Physical basis and biological mechanisms of gold nanoparticle radiosensitization. *Nanoscale* **4**, 4830-4838 (2012).
17  Huang, X., El-Sayed, I. H., Qian, W. & El-Sayed, M. A. Cancer cell imaging and photothermal therapy in the near-infrared region by using gold nanorods. *J. Am. Chem. Soc.* **128**, 2115-2120 (2006).
18  Hainfeld, J. F., Slatkin, D. N. & Smilowitz, H. M. The use of gold nanoparticles to enhance radiotherapy in mice. *Phys. Med. Biol.* **49**, N309 (2004).
19  Chithrani, D. B. *et al.* Gold nanoparticles as radiation sensitizers in cancer therapy. *Radiat. Res.* **173**, 719-728 (2010).
20  Rahman, W. N. *et al.* Enhancement of radiation effects by gold nanoparticles for superficial radiation therapy. *Nanomed-Nanotechnol.* **5**, 136-142 (2009).
21  Roa, W. *et al.* Gold nanoparticle sensitize radiotherapy of prostate cancer cells by regulation of the cell cycle. *Nanotechnology* **20**, 375101 (2009).
22  Hainfeld, J. F., Dilmanian, F. A., Slatkin, D. N. & Smilowitz, H. M. Radiotherapy enhancement with gold nanoparticles. *J. Pharm. Pharmacol.* **60**, 977-985 (2008).
23  Lechtman, E. *et al.* Implications on clinical scenario of gold nanoparticle radiosensitization in regards to photon energy, nanoparticle size, concentration and location. *Phys. Med. Biol.* **56**, 4631 (2011).
24  McMahon, S. J. *et al.* Biological consequences of nanoscale energy deposition near irradiated heavy atom nanoparticles. *Sci. Rep.* **1**, 1-9 (2011).
25  McMahon, S. J., Mendenhall, M. H., Jain, S. & Currell, F. Radiotherapy in the presence of contrast agents: a general figure of merit and its application to gold nanoparticles. *Phys. Med. Biol.* **53**, 5635 (2008).
26  Zhang, G. *et al.* Influence of anchoring ligands and particle size on the colloidal stability and in vivo biodistribution of polyethylene glycol-coated gold nanoparticles in tumor-xenografted mice. *Biomaterials* **30**, 1928-1936 (2009).
27  von Maltzahn, G. *et al.* Computationally guided photothermal tumor therapy using long-circulating gold nanorod antennas. *Cancer Res.* **69**, 3892-3900 (2009).
28  Huo, S. *et al.* Superior penetration and retention behavior of 50 nm gold nanoparticles in tumors. *Cancer Res.* **73**, 319-330 (2013).
29  Huang, X. *et al.* A reexamination of active and rassive tumor targeting by using rod-shaped gold nanocrystals and covalently conjugated peptide ligands. *ACS Nano* **4**, 5887-5896 (2010).
30  Chou, L. Y. T. & Chan, W. C. W. Fluorescence-tagged gold nanoparticles for rapidly characterizing the size-dependent biodistribution in tumor models. *Adv. Healthcare Mater.* **1**, 714-721 (2012).
31  Choi, C. H. J., Alabi, C. A., Webster, P. & Davis, M. E. Mechanism of active targeting in solid tumors with transferrin-containing gold nanoparticles. *Proc. Natl. Acad. Sci. U.S.A.* **107**, 1235-1240 (2010).
32  Ma, X. *et al.* Gold nanoparticles induce autophagosome accumulation through size-dependent nanoparticle uptake and lysosome impairment. *ACS Nano* **5**, 8629-8639 (2011).
33  Zhang, X. D. *et al.* Ultrasmall $Au_{10-12}(SG)_{10-12}$ nanomolecules for high tumor specificity and cancer radiotherapy. *Adv. Mater.* **26**, 4565–4568 (2014).
34  Choi, H. S. *et al.* Renal clearance of quantum dots. *Nat. Biotech.* **25**, 1165-1170 (2007).
35  Setyawati, M. *et al.* Titanium dioxide nanomaterials cause endothelial cell leakiness by disrupting the homophilic interaction of VE–cadherin. *Nat. Commun.* **4**, 1673 (2013).
36  Tay, C. Y. *et al.* Nanoparticles strengthen intracellular tension and retard cellular migration. *Nano lett.* **14**, 83-88 (2013).
37  Zhang, X.-D. *et al.* Size-dependent radiosensitization of PEG-coated gold nanoparticles for cancer radiation therapy. *Biomaterials* **33**, 6408-6419 (2012).
38  Lundqvist, M. *et al.* Nanoparticle size and surface properties determine the protein corona with possible implications for biological impacts, *Proc. Natl. Acad. Sci. U.S.A.* **23**, 14265–14270 (2008).
39  Luo, Z., Zheng, K. & Xie, J. Engineering ultrasmall water-soluble gold and silver nanoclusters for biomedical applications. *Chem. Commun.* **50**, 5143-5155 (2014).
40  Luo, Z. *et al.* From aggregation-induced emission of Au(I)–thiolate complexes to ultrabright Au(0)@Au(I)–thiolate core–shell nanoclusters. *J. Am. Chem. Soc.* **134**, 16662-16670 (2012).
41  Yu, Y. *et al.* Identification of a highly luminescent $Au_{22}(SG)_{18}$ nanocluster. *J. Am. Chem. Soc.* **136**, 1246-1249 (2014).
42  Luo, Z. *et al.* Toward understanding the growth mechanism: tracing all stable intermediate species from reduction of Au (I)–thiolate complexes to evolution of $Au_{25}$ nanoclusters. *J. Am. Chem. Soc.* **136**, 10577-10580 (2014).
43  Yuan, X. *et al.* Balancing the rate of cluster growth and etching for gram‐scale synthesis of thiolate‐protected $Au_{25}$ nanoclusters with atomic precision. *Angew. Chem. Int. Ed.* **53**, 4623-4627 (2014).
44  Jin, R. Quantum sized, thiolate-protected gold nanoclusters. *Nanoscale* **2**, 343-362 (2010).
45  Dou, X. *et al.* Lighting up thiolated Au@Ag nanoclusters via aggregation-induced emission. *Nanoscale* **6**, 157-161 (2014).
46  Zhang, X.-D. *et al.* Enhanced tumor accumulation of sub-2 nm gold nanoclusters for cancer radiation therapy. *Adv. Healthcare Mater.* **3**, 133-141 (2014).
47  Liu, J. *et al.* Passive tumor targeting of renal-clearable luminescent gold nanoparticles: long tumor retention and fast normal tissue clearance. *J. Am. Chem. Soc.* **135**, 4978-4981 (2013).
48  Zhou, C., Long, M., Qin, Y., Sun, X. & Zheng, J. Luminescent gold nanoparticles with efficient renal clearance. *Angew. Chem. Int. Ed.* **50**, 3168-3172 (2011).
49  Zhang, X.-D. *et al.* In vivo renal clearance, biodistribution, toxicity of gold nanoclusters. *Biomaterials* **33**, 4628-4638 (2012).
50  Liu, Z. *et al.* In vivo biodistribution and highly efficient tumour targeting of carbon nanotubes in mice. *Nat. Nanotechnol.* **2**, 47-52 (2007).
51  Yang, K. *et al.* Graphene in mice: ultrahigh in vivo tumor uptake and efficient photothermal therapy. *Nano letters* **10**, 3318-3323 (2010).
52  Tay, C. Y., Setyawati, M. I., Xie, J., Parak, W. J. & Leong, D. T. Back to basics: exploiting the innate physico‐chemical characteristics of nanomaterials for biomedical applications. *Adv. Funct. Mater.* **24**, 5936–5955 (2014).

53  Vinluan III, R. D. *et al.* Glutathione-coated luminescent gold nanoparticles: a surface ligand for minimizing serum protein adsorption. *ACS Appl. Mater. Inter.* **6**, 11829-11833 (2014).
54  Zhang, X.-D. *et al.* Toxicologic effects of gold nanoparticles in vivo by different administration routes. *Int. J. Nanomed.* **5**, 771-781 (2010).
55  Zhang, X.-D. *et al.* Size-dependent in vivo toxicity of PEG-coated gold nanoparticles. *Int. J. Nanomed.* **6**, 2071-2081 (2011).
56  Chen, J. *et al.* Sex differences in the toxicity of polyethylene glycol-coated gold nanoparticles in mice. *Int. J. Nanomed.* **8**, 2409-2419 (2013).
57  Pinzani, V. *et al.* Cisplatin-induced renal toxicity and toxicity-modulating strategies: a review. *Cancer Chemother. Pharmacol.* **35**, 1-9 (1994).
58  Yu, Y., Luo, Z., Yu, Y., Lee, J. Y. & Xie, J. Observation of cluster size growth in CO-directed synthesis of $Au_{25}(SR)_{18}$ nanoclusters. *ACS Nano* **6**, 7920-7927 (2012).

**Figure Captions**

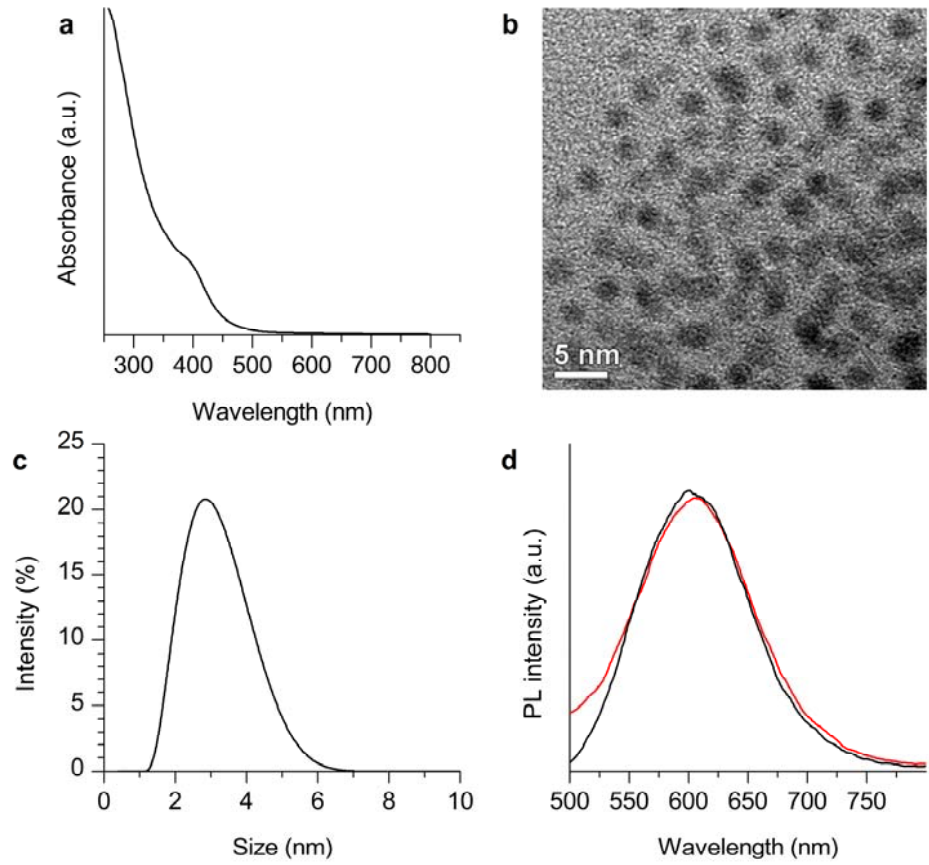

**Figure 1.** (a) UV-vis absorption spectrum, (b) TEM image, and (c) hydrodynamic diameter (measured by dynamic light scattering) of the as-prepared $Au_{29-43}(SG)_{27-37}$ NCs. (d) Photoluminescence spectra ($\lambda_{ex}$ = 365 nm) of $Au_{29-43}(SG)_{27-37}$ NCs (black line) and the mixture of $Au_{29-43}(SG)_{27-37}$ NCs and blood plasma (at 24 h after mixing, red line).

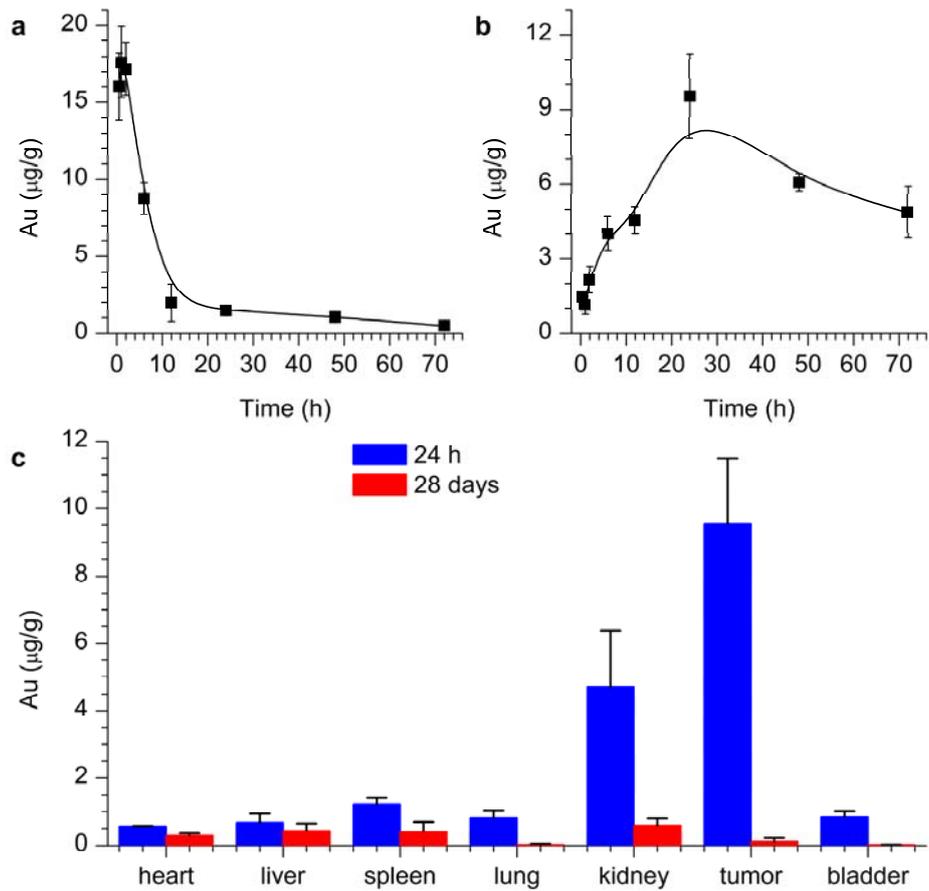

**Figure 2**. (a) *In vivo* blood concentration studies of Au$_{29-43}$(SG)$_{27-37}$ NCs. (b) Tumor uptake of Au$_{29-43}$(SG)$_{27-37}$ NCs after different time injection. (c) Biodistribution of Au$_{29-43}$(SG)$_{27-37}$ NCs after 24 h and 28 days p.i.

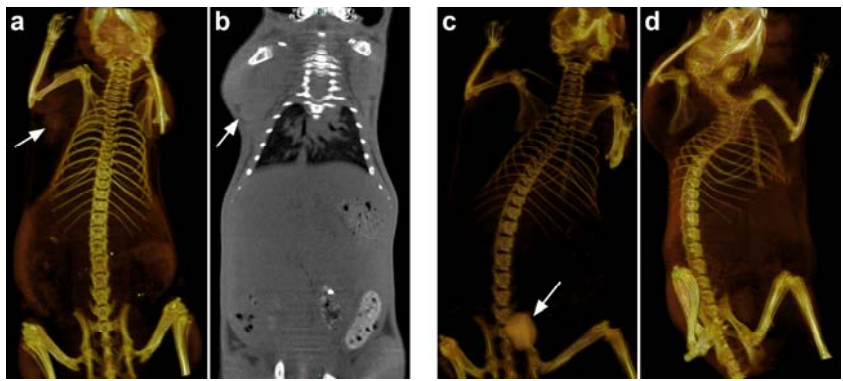

**Figure 3**. Small animal X-ray computed tomography (a) three-dimensional and (b) two-dimensional imaging of Au$_{29-43}$(SG)$_{27-37}$ NCs at 6 h p.i. using U14 tumor bearing mice. Renal clearance of Au$_{29-43}$(SG)$_{27-37}$ NCs at the time point of (c) 1 h and (d) 24 h p.i. using nude mice without tumor.

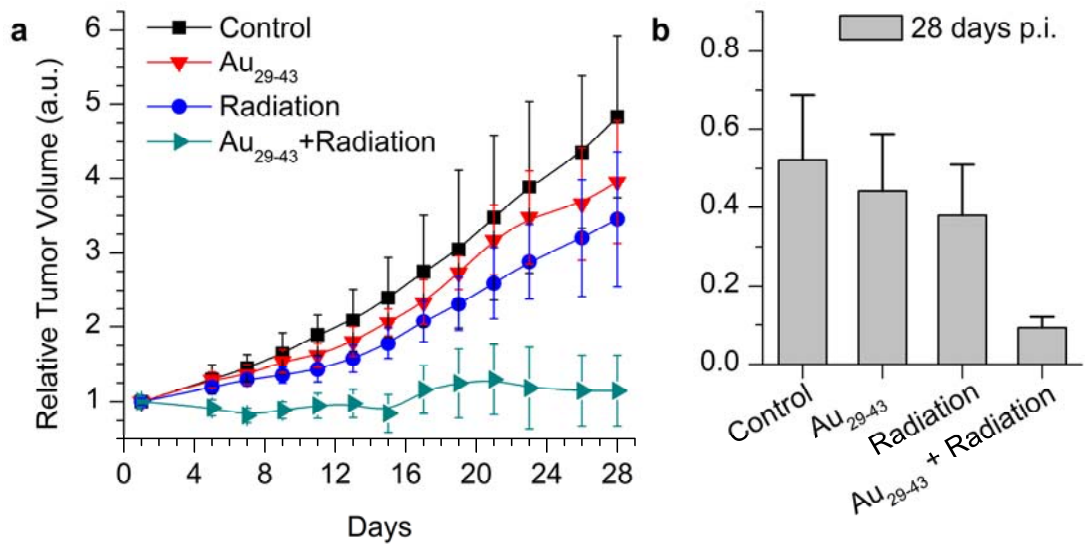

**Figure 4**. Time-course studies of tumor (a) volume and (b) weight of mice treated with $Au_{29-43}(SG)_{27-37}$ NCs at the concentration of 5.9 mg-Au/kg-body.

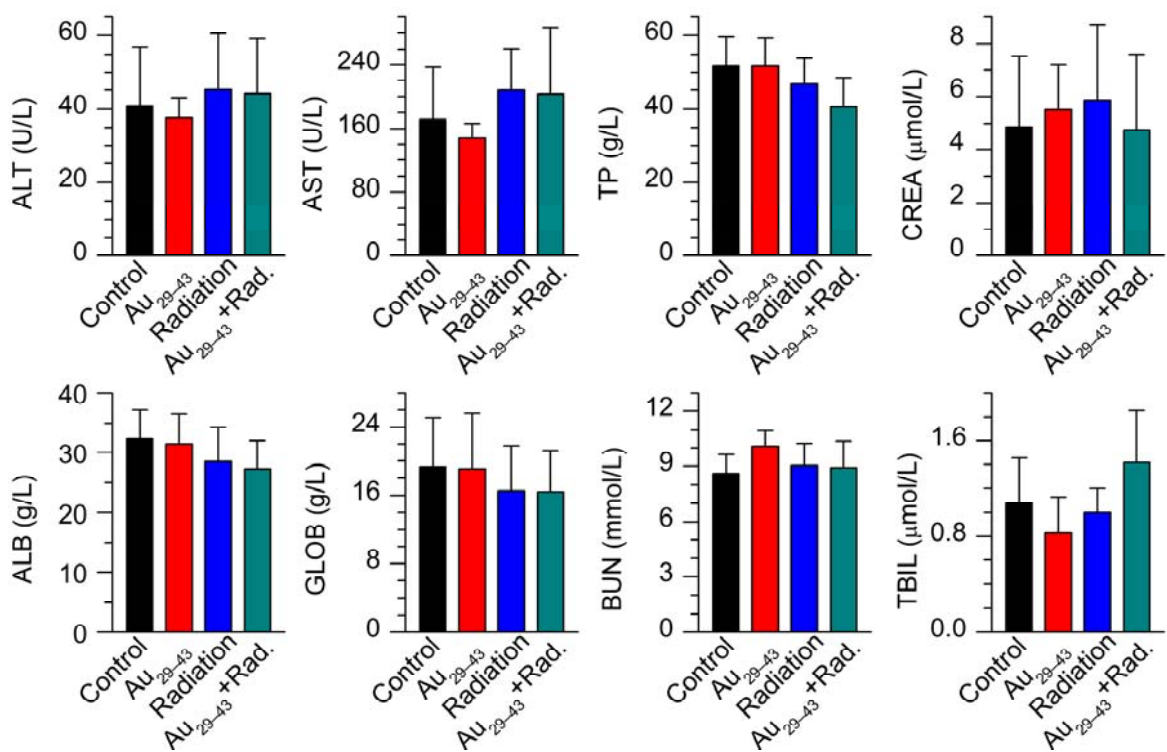

**Figure 5.** Blood biochemistry analysis of mice treated with $Au_{29-43}(SG)_{27-37}$ NCs at 28 days p.i.. The results show mean and standard deviation of alanine aminotransferase (ALT), aspartate aminotransferase (AST), total protein (TP), albumin (ALB), blood urea nitrogen (BUN), creatinine (CREA), globulin (GOLB), and total bilirubin (TB).

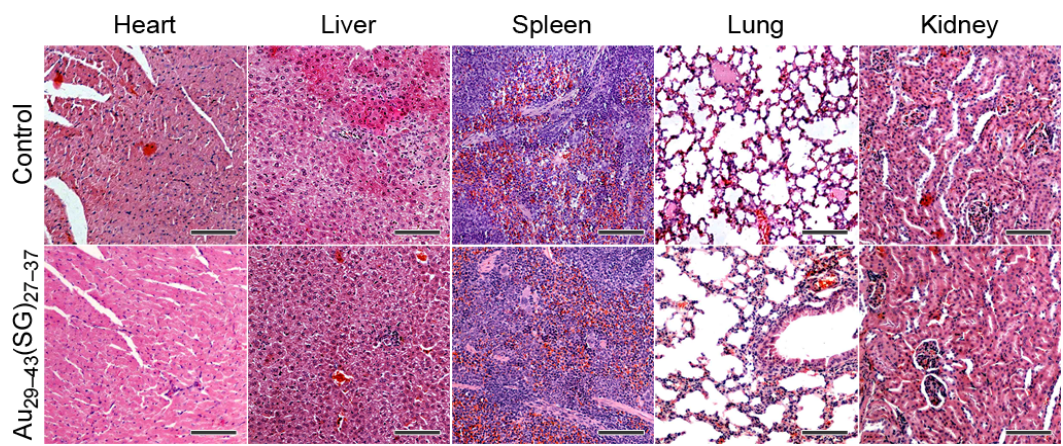

**Figure 6.** Pathological data from the heart, liver, spleen, lung, and kidney of mice treated with $Au_{29-43}(SG)_{27-37}$ NCs at the concentration of 5.9 mg-Au/kg-body. Scale bars, 100 μm.

**Acknowledgment**

This work was supported by the National Natural Science Foundation of China (Grant No.81471786 and 11304220), Natural Science Foundation of Tianjin (Grant No. 13JCQNJC13500) and Foundation of Union New Star, CAMS (No.1256). Part of this work was supported by the Ministry of Education, Singapore, under grant R-279-000-409-112.

**Author contributions**

X. Z., Z. L., J.X., and M. G. conceived the project and designed the experiments. J. C., Z. L., and X.S., S. S., X. Y., and X. Z. performed the experiments. Z. L., H. W., and X. Y. synthesized the materials and J. C., X. S, L. Z., K. G., Y. S., and S. S. performed the *in vivo* experiment. X. Z., Z. L., S. F., D. T. L., and J. X. analyzed the data and co-wrote the paper. All authors discussed the results and commented on the manuscript.